\documentclass[10pt]{article}
\usepackage{graphicx}

\def\Title#1{\begin{center} {\Large #1 } \end{center}}
\def\Author#1{\begin{center}{ \sc #1} \end{center}}
\def\Address#1{\begin{center}{ \it #1} \end{center}}

\newcommand\pubblock{\rightline{\begin{tabular}{l} Proceedings of the Second Annual LHCP\\ \pubnumber\\
         \pubdate  \end{tabular}}}

\newenvironment{Abstract}{\begin{quotation} \begin{center} 
             \large ABSTRACT \end{center}\bigskip 
      \begin{center}\begin{large}}{\end{large}\end{center} \end{quotation}}

\newenvironment{Presented}{\begin{quotation} \begin{center} 
             PRESENTED AT\end{center}\bigskip 
      \begin{center}\begin{large}}{\end{large}\end{center} \end{quotation}}

\def\beq{\begin{equation}}
\def\eeq#1{\label{#1}\end{equation}}
\def\eeqn{\end{equation}}

\def\beqa{\begin{eqnarray}}
\def\eeqa#1{\label{#1}\end{eqnarray}}
\def\eeqan{\end{eqnarray}}

\let\bar=\overbar

\def\Dslash{\not{\hbox{\kern-4pt $D$}}}
\def\dslash{\not{\hbox{\kern-2pt $\del$}}}

\def\msb{{\bar{\ssstyle M \kern -1pt S}}}

\usepackage{color}

 \newcommand\pubnumber{ }

\newcommand\pubdate{\today}

\def\affiliation{
On behalf of the ALICE Collaboration, \\
Department of Physics \\
Yale University, New Haven, CT 06511, U.S.A. }

\begin{document}

\large
\begin{titlepage}
\pubblock

\vfill
\Title{  Jet Production in p-Pb Collisions  }
\vfill

\Author{Megan Connors  }
\Address{\affiliation}
\vfill
\begin{Abstract}

One of the major results from the study of high energy heavy ion
collisions is the observation of jet quenching. The suppression of the
number of jets observed in heavy ion collisions relative to pp
collisions at the same energy scaled by the number of binary
collisions, is attributed to partonic energy loss in the quark gluon
plasma (QGP). However, cold nuclear matter effects due to the presence
of a nucleus in the initial state could also influence this
measurement. To disentangle these effects p-Pb collisions are studied,
where QGP formation is not expected to occur and only cold nuclear
matter effects are present. In addition to being an important baseline
for understanding jet quenching, jets in p-Pb collisions may also be
used to provide constraints on the nuclear parton distribution
functions. Fully reconstructed jets measured using the ALICE tracking
system and electro-magnetic calorimeter in p-Pb collisions at
$\sqrt{s_{NN}}=5.02$ TeV are reported. In addition to the spectra,
studies of the jet fragmentation behavior in p-Pb collisions are also presented.

\end{Abstract}
\vfill

\begin{Presented}
The Second Annual Conference\\
 on Large Hadron Collider Physics \\
Columbia University, New York, U.S.A \\ 
June 2-7, 2014
\end{Presented}
\vfill
\end{titlepage}
\def\thefootnote{\fnsymbol{footnote}}
\setcounter{footnote}{0}
%

\normalsize 


\section{Introduction}

Heavy ions collisions at high energies produce a state of matter
called the Quark Gluon Plasma (QGP). Jets are an excellent probe of the QGP, since they originate from hard scatterings early
in the collision. As the produced parton traverses the
QGP, it loses energy. The result is
jet quenching, which has been observed and is quantified by comparing the fully
corrected jet spectrum measured in central Pb-Pb collisions to that measured in pp
scaled by the average number of binary collisions, $\langle N_{\rm
  coll}\rangle$ \cite{alicePbPbjets}. However, since the parton experiences all stages of the
collision, cold nuclear matter (CNM)
effects could also influence this
measurement. Measuring the jet cross section in p-Pb collisions,
which experience CNM effects without the
QGP, is critical for disentangling initial and final state effects on
the observed Pb-Pb jet spectra. 

In addition to modifications of the total jet production, CNM could
also influence the observed jet fragmentation. To study the fragmentation properties of the jet, 
the spectra for jets reconstructed at different radii are compared. A modification to the jet substructure
would be observed if their ratio in p-Pb
differs from the same ratio measured in pp
collisions. However to further quantify the fragmentation behavior, the $j_{T}$, which is the vector between the charged track and
the jet axis, is measured. The $j_{T}$ distribution in p-Pb is then compared to PYTHIA. 

Another important observation from heavy ion collisions has been the
increased baryon to meson ratio for high multiplicity Pb-Pb collisions compared
to low multiplicity events \cite{lamkPbPb}. A similar behavior has been observed in
p-Pb collisions \cite{lamkpPb}. For both systems, the $\Lambda /K^0_{s}$
ratio for $p_{T}>2$ GeV/$c$ is higher for high multiplicity events
than for low multiplicity events. In Pb-Pb, this observation has been
attributed to radial flow. The applicability of hydrodynamics to
smaller systems such as p-Pb is still under investigation. While this
modification has been observed, it is unclear whether the composition
of the jet is being modified or if it arises from soft (small $Q^{2}$)
processes in the underlying event.

\section{Analysis Details}
\label{analysis}

Three independent analyses studing jets in the p-Pb collisions are presented: the fully reconstructed jet
spectrum, the $j_{T}$ distribution and the measurement of V$^{0}$
particles in charged jets. Since the methods used in these analyses
overlap, a general discussion of the analysis details is
presented. Cases where there are differences in the procedures are
noted. 

The presented analyses all use data collected by the ALICE experiment
during the 5.02 TeV p-Pb LHC run from 2013. The jet spectrum and
V$^{0}$ study use minimum bias p-Pb events corresponding to an
integrated luminosity of 51$\mu b^{-1}$ while the $j_{T}$ study also
incorporates events triggered by the electromagentic calorimeter
(EMCal). Event multiplicity classes are determined using the VZERO-A
detector which covers the psuedorapidity, $2.8 < \eta <5.1$ in the Pb going
direction. Charged tracks are measured in the ALICE central tracking
system, which consists of the Time Projection Chamber (TPC) and a silicon Inner
Tracking System (ITS). To account for the neutral energy of the jet,
clusters are measured in the EMCal, which covers 110$^{\circ}$ in
azimuth and $|\eta|<0.7$. Clusters are corrected for energy
deposited in the EMCal by charged tracks \cite{aliceppjets}.

\subsection{Jet Reconstruction and Underlying Event Correction}
For full jet reconstruction
EMCal clusters with $E_{T}>300$ MeV and
charged tracks with $p_{T} > 150$ MeV/$c$ are input to the jet finding
algorithm. Note that the jets in the V$^{0}$ study only include the charged
tracks. 
Jets are reconstructed with the anti-$k_{T}$ jet finding
algorithm with various resolution parameters, $R$ using the
FastJet package \cite{fastjet}. To ensure the jet is fully within the
acceptance, the jet axis must be at least $R$ away from
the edge of the detector. The V$^{0}$ measurement restricts the jets to $|\eta|<0.75-R$.

Energy from the underlying event is also clustered into the
jets by the algorithm and must be subtracted from the total raw jet
energy. An average energy density, $\rho$, is determined on an event-by-event
basis and then subtracted from each jet in the event according to
$p_{T,jet}^{reco}=p_{T,jet}^{raw}-\rho \times A_{jet}$ where $A_{jet}$
is the area of the reconstructed jet.  For
the full jet analyses, the charged track $p_{T}$ density, $\rho^{ch}$, which
is measured in full azimuth is scaled using a scale factor that is
determined from measured data to include electromagnetic contributions, as done in the Pb-Pb jet spectra
analysis \cite{alicePbPbjets}.  The charged track
background density, $\rho^{ch}$, is determined by the median occupancy
method, a slightly modified implementation of the method
presented in \cite{cmsrho}. The median occupancy method,
defined by
\begin{equation}
\label{eq:rho}
\rho^{ch}=\rm{median}\left \{\frac{p_{T}^{i}}{A_{i}} \right \} \times C,
\end{equation}
is determined by running the $k_{T}$ algorithm over all tracks plus ``ghost
particles'' (fake particles with negligible momentum used by FastJet
for the jet area calculation) in the event \cite{fastjet}. Before determining the
median of the physical jets, any $k_{T}$ jets overlapping with signal jets are
excluded. The median is scaled by an occupancy correction
factor, $C$, accounting for the emptiness of
the p-Pb event. $C=A_{\rm{phys}}/A_{\rm{tot}}$, where $A_{\rm{phys}}$ is the area of all physical jets and
$A_{tot}$ is the area of all jets, including jets comprised of
ghost particles only.

\subsection{V$^{0}$ Candidate Selection and Underlying Event Correction}

The $\Lambda$ and $K^{0}_{s}$ candidates are reconstructed via their
hadronic decay channels, $\Lambda\rightarrow p\pi^{-}$ and $K^{0}_{s}
\rightarrow \pi^{+}\pi^{-}$. The decay daughters are identificed in
the TPC according to their specific ionization, d$E$/d$x$. A
fiducial cut was applied requiring all V$^{0}$ particles satisfy
$|\eta_{\rm{V}^{0}}<0.75|$. Additional
details can be found in \cite{V0QM}. 

V$^{0}$ candidates are considered
to be part of a jet if the distance between the candidate and the jet
axis is less than the resolution parameter, $R$. 
The number of V$^{0}$
candidates within
the jet cone is corrected for the underlying event. To estimate the
contribution from V$^{0}$ particles not associated with the hard
scattering, V$^{0}$ particles are measured outside the jet cone and in
non-jet events. The difference in the spectra from these two different
selection criteria estimates the systematic uncertainty on the
background determination. Feed-down of $\Lambda$ from $\Xi$ is corrected
using the feed-down fraction of inclusive $\Lambda$ determined from
the data.



\subsection{Unfolding}
\label{uf}

After background subtraction, the jet spectra must be corrected for detector effects and fluctuations in the underlying event. The effect of the detector on the spectra is determined by
passing PYTHIA events at $\sqrt{s} =$ 5.02 TeV through a GEANT
simulation of the ALICE detector.  Jets reconstructed at the detector level are geometrically matched to the closest particle
level jet. A
2-dimensional histogram or response matrix (RM) maps
detector level jet $p_{T}$ to particle level jet $p_{T}$. More details
are available in \cite{myQM}.

The underlying event energy density, is determined on an event-by-event
basis. However, the background energy density fluctuates
within each event. These fluctuations can be quantified by
measuring the $\delta p_{T}$ distribution using the method of
Random Cones (RC) according to
\begin{equation}
\delta p_{T}=p_{T}^{RC}-\pi R_{RC}^2 \times \rho, 
\end{equation}
where $p_{T}^{RC}$, is the total momentum within a cone of
radius, $R_{RC}$, placed randomly in the event.  

The final RM, the multiplication of the detector RM and the $\delta p_{T}$ distribution, is input to the unfolding algorithm. The singular value decomposition (SVD) algorithm was chosen as the
default for unfolding the spectrum \cite{svd}. A bin-by-bin correction procedure was
applied for the $j_{T}$ analysis. For the jet spectra measurement, the bin-by-bin method was also in good agreement with the other unfolding algorithms. 

\section{Results}

\begin{figure}[hbtp]
\centerline
{
\includegraphics[width=0.35\textwidth]{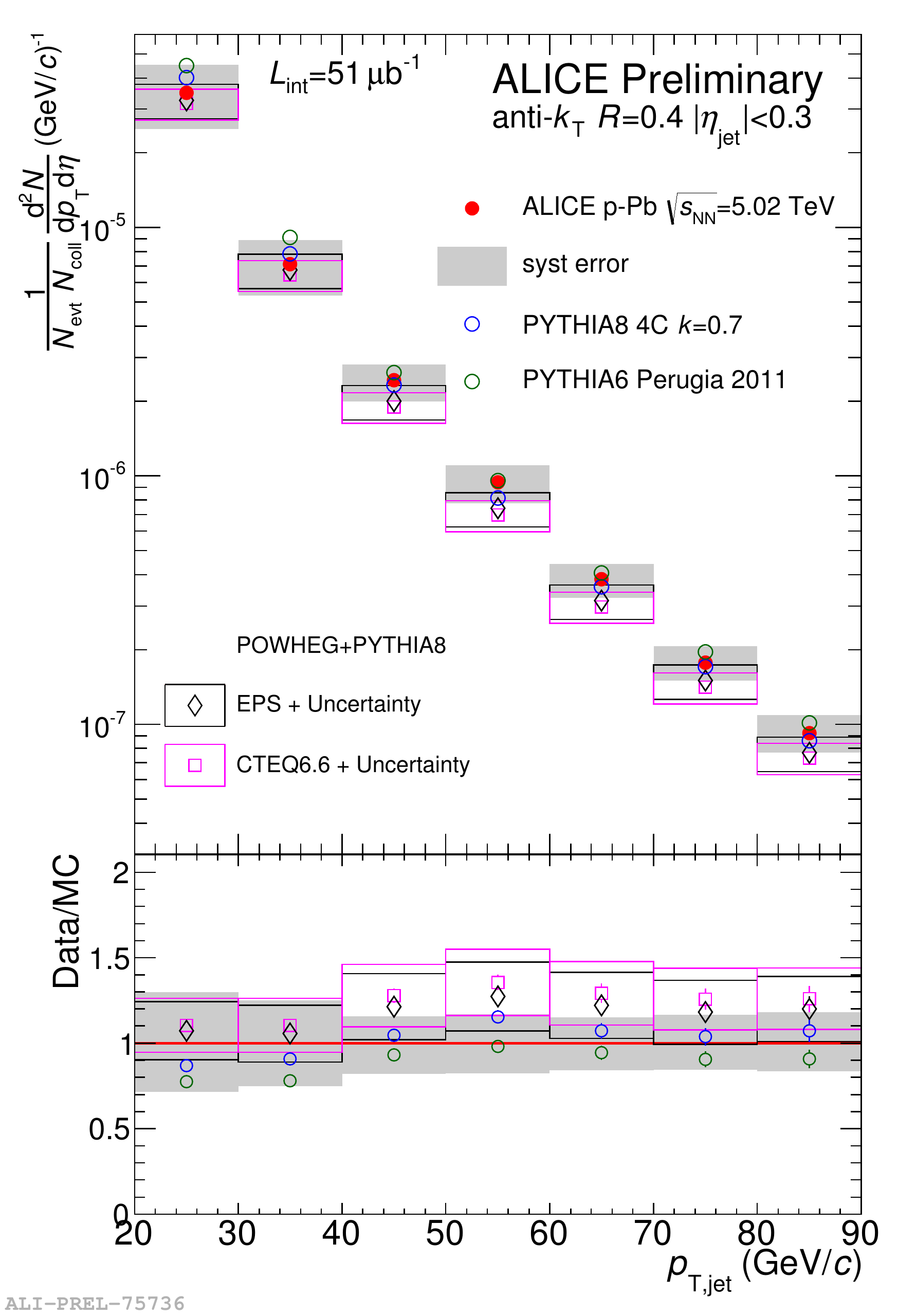}
\includegraphics[width=0.35\textwidth]{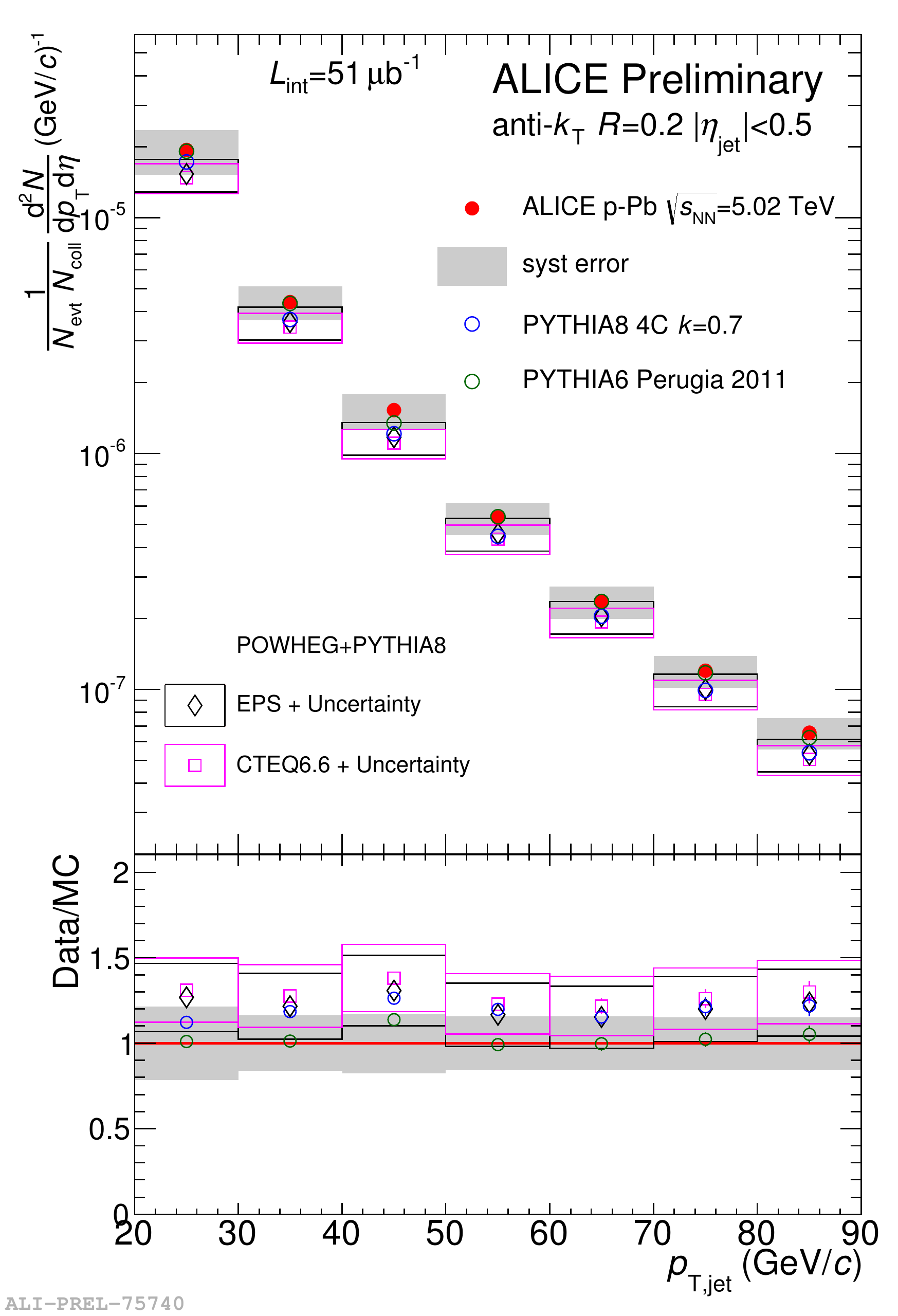}
  }
  \caption{Fully corrected p-Pb jet spectrum for $R$ = 0.4 and $R$ = 0.2
    scaled by $\langle N_{\rm coll}\rangle$
    compared to PYTHIA and POWHEG simulations at 5.02 TeV.}
\label{fig:spectra}
\end{figure}

The unfolded spectra for $R$ = 0.4 and $R$ = 0.2 are shown in
Figure \ref{fig:spectra}. The spectra are normalized per number of
binary collision, $\langle N_{\rm coll}\rangle$, to make a direct comparison to the pp
references. Since no data exists on pp collisions at
$\sqrt{s}$ = 5.02 TeV, we compare the p-Pb data to Monte Carlo simulations. The plot
includes PYTHIA8, PYTHIA6 and POWHEG using 2 different parton
distribution functions (PDF). The POWHEG calculations include
uncertainties on the factorization and renormalization scales (13\%) and the uncertainty in the PDF
(6\% for CTEQ and 9\% for EPS). The ratios between the data and the
different models are all consistent with one, which would suggest there are no
cold nuclear matter effects to the jet spectrum. However, the spread and uncertainty from these
different references is significant and demonstrates the need for a data
reference to better quantify the CNM effects or lack of effects on the
jet spectrum. Despite this uncertainty,
the p-Pb results clearly show that the strong suppression observed in the Pb-Pb is not
purely due to initial state effects, but is rather a result of
energy loss in the produced medium. 

\begin{figure}[htbp]
\centerline
{
\includegraphics[width=0.5\textwidth]{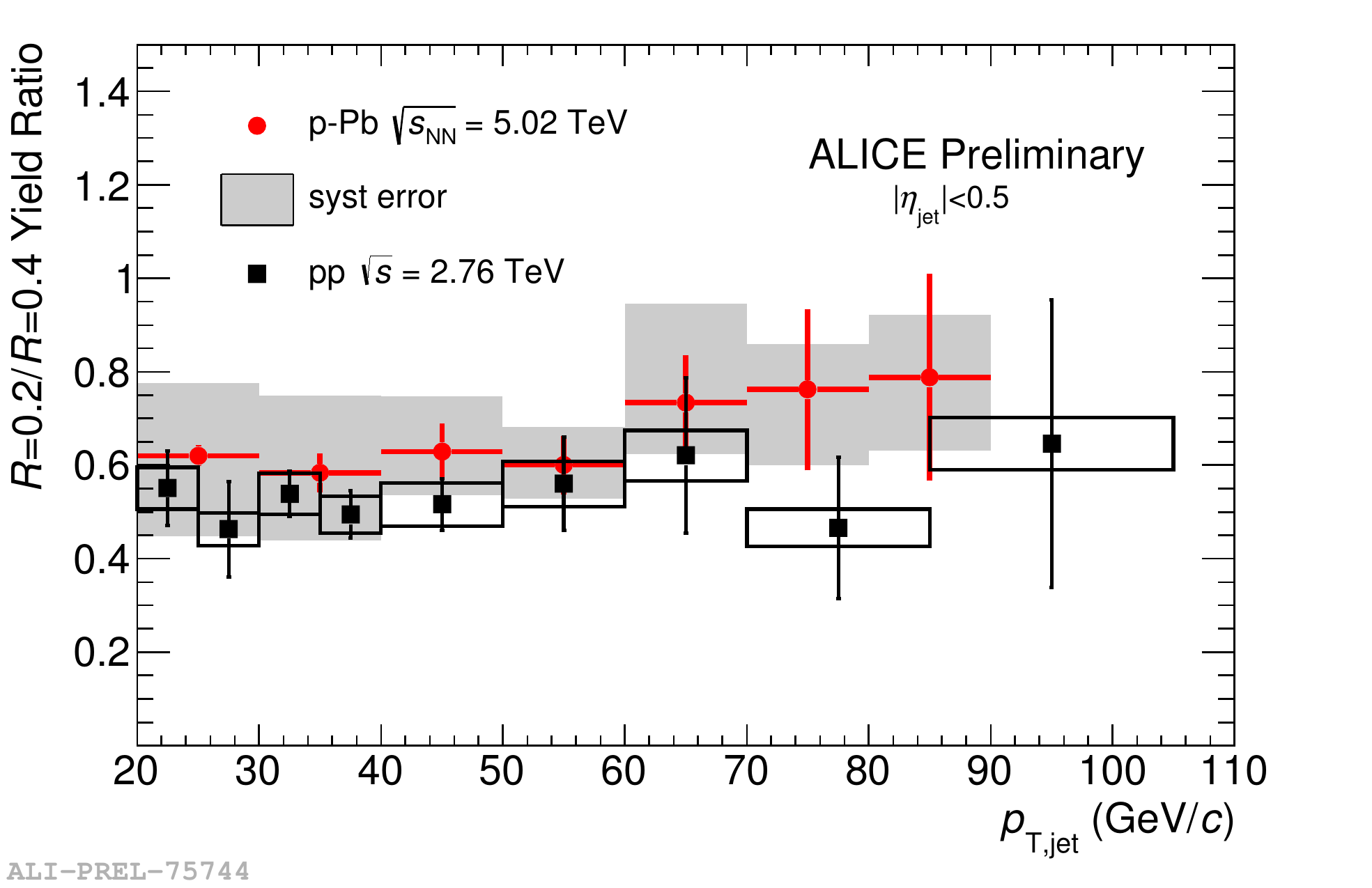}
  }
  \caption{Cross-section ratio between $R$ = 0.2 jet spectrum and $R$ =
    0.4
    jet spectrum for fully reconstructed jets in 5.02 TeV p-Pb
    collisions (red circles) and 2.76 TeV pp collisions (black squares).}
\label{fig:JSR}
\end{figure}

Although the jet spectra appear unmodified, one may question whether
the fragmentation could still be altered in p-Pb
collisions. The ratio of the spectra measured with different $R$
provides insight to fragmentation behavior of the jet. Figure
\ref{fig:JSR} shows the ratio of the $R$ = 0.2 spectrum to the $R$ = 0.4
spectrum for 5.02 TeV p-Pb (red circles) and for 2.76 TeV pp (black
squares) collisions.  The agreement between the two systems
suggests that the fragmentation behavior for jets in p-Pb is very
similar to that in pp collisions. 

\begin{figure}[htp]
\centerline
{
\includegraphics[width=0.65\textwidth]{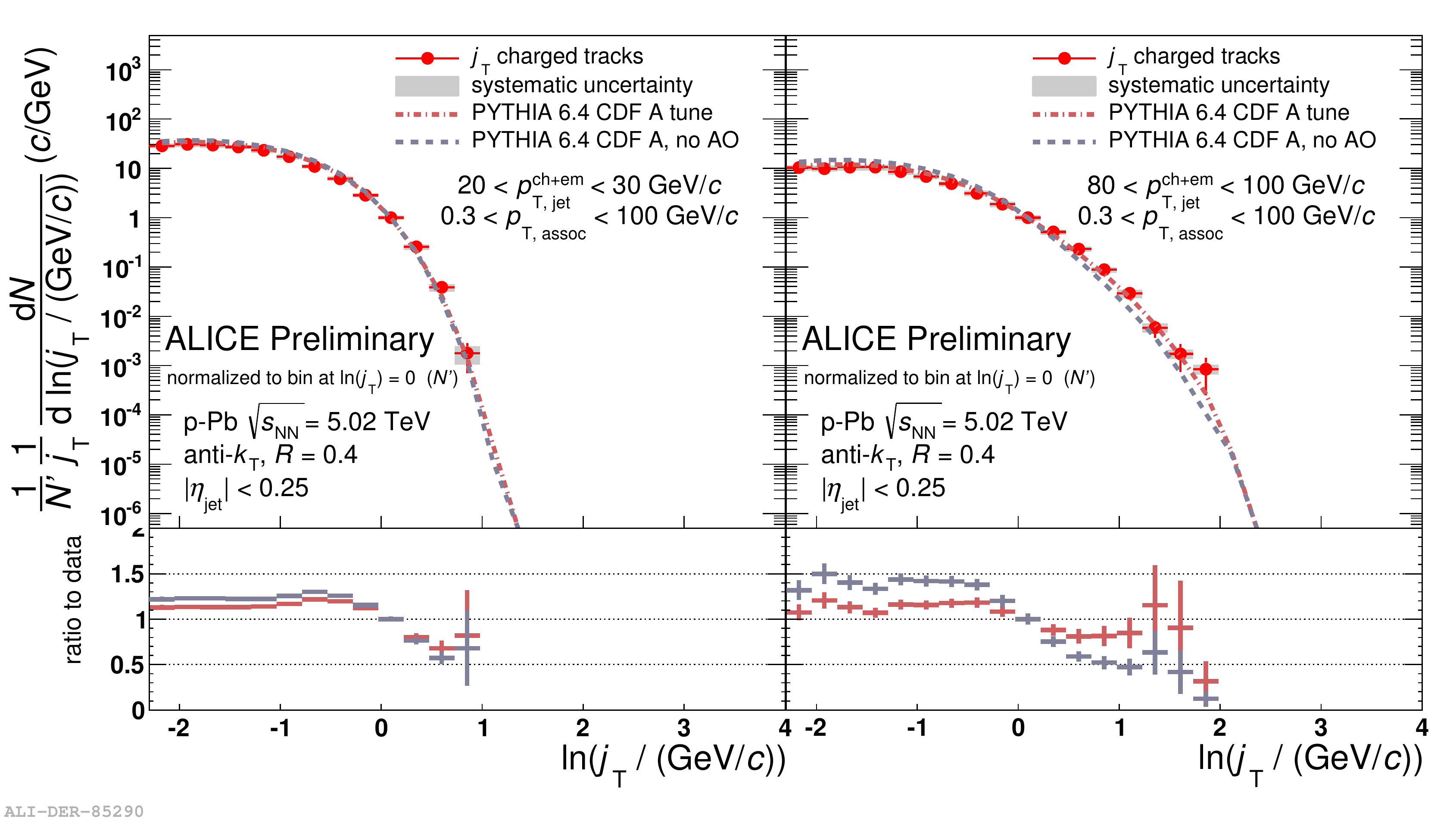}
 }
  \caption{$j_{T}$ for anti-k$_{T}$ $R$ = 0.4 jets with $20 < p_{T} < 30$
    GeV/$c$ (left) and $80 < p_{T} < 100$ GeV/$c$ (right)
    compared to PYTHIA 6.4 CDF A tune with (red) and
    without (gray)
    angular ordering (AO).}
\label{fig:jt}
\end{figure}

A more differential approach for probing the fragmentation properties
of the jet is by measuring the $j_{T}$ distribution where $j_{T}$ is
the momentum component perpendicular to the jet axis. The measured
distribution is shown in Figure \ref{fig:jt} for two different jet
momentum ranges. The distributions
are compared to PYTHIA 6.4 using the CDF A tune with angular
ordering on (red dashed line) and off (gray dashed line). The shape of
the distribution appears to be well reproduced by the simulation. The ratio
between the PYTHIA and data plotted in
the lower panels of Figure \ref{fig:jt} show better agreement when the
angular ordering is turned on.

\begin{figure}[htp]
\centerline
{
\includegraphics[width=0.75\textwidth]{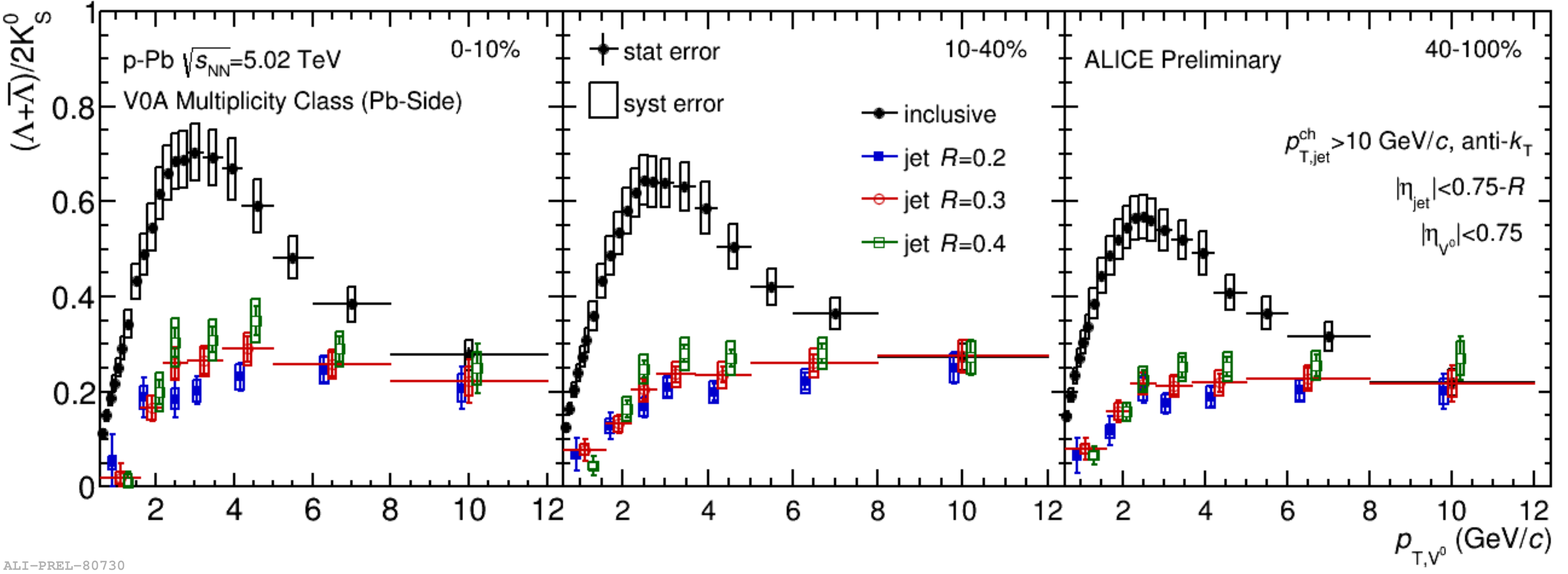}
  }
  \caption{The $\Lambda/K^{0}_{s}$ ratio within charged jets with  $p^{ch}_{T,jet} > 10$ GeV/$c$ reconstructed for different $R$ compared
    to the inclusive ratio in black for
    multiplicity classes 0-10\% (left), 10-40\% (middle) and 40-80\%
    (right).}
\label{fig:PartCompMultDep}
\end{figure}

Finally, the $\Lambda/K^{0}_{s}$ ratio within a jet is measured for charged
jets with $p^{ch}_{T,jet} > 10$ GeV/$c$ and plotted in
Figure \ref{fig:PartCompMultDep}. The plot includes jets reconstructed with the
anti-$k_{T}$ algorithm for different resolution parameters, R=0.2
(solid blue squares), R=0.3 (red open circles) and R=0.4 (green open squares). Little to no dependence on
the resolution parameter of the jet is observed. For comparison
the inclusive ratio is also plotted as solid black circles in Figure
\ref{fig:PartCompMultDep}. The inclusive ratio is clearly larger than the ratio
observed within the jet.  The panels show the
ratio measured in three different multiplicity classes. While the inclusive ratio appears to decrease as a
function of multiplicity, the ratio measured in the jets stays
constant. 

\begin{figure}[htp]
\centerline
{
\includegraphics[width=0.58\textwidth]{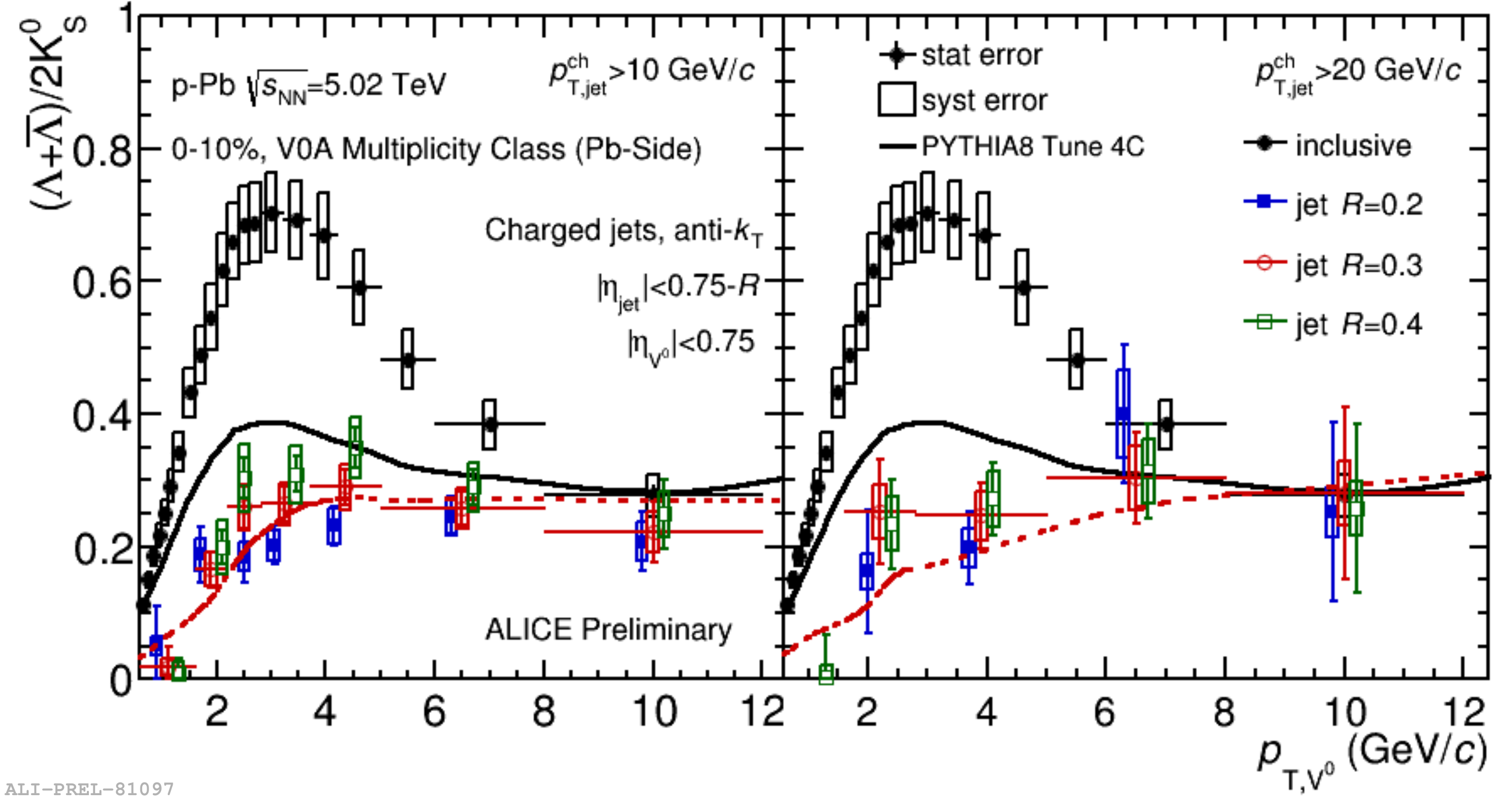}
  }
  \caption{The inclusive $\Lambda/K^{0}_{s}$ ratio and the
    ratio within jets reconstructed for different R compared to PYTHIA
    curves for the inclusive ratio (black solid line) and ratio within
    R=0.3 jets (red dashed
    line) for charged jets with $p^{ch}_{T,jet} > 10$
    GeV/$c$ (left) and $p^{ch}_{T,jet} > 20$ GeV/$c$.}
\label{fig:ParticleComposition}
\end{figure}

In Figure \ref{fig:ParticleComposition}, the data are compared to
results from 
PYTHIA8 with Tune 4C. The inclusive $\Lambda/K^{0}_{s}$ ratio is also
higher than the ratio within the jet for PYTHIA. However, the ratio
within the jet agrees with the data while the data shows a clear
enhancement over PYTHIA for the inclusive measurement. These results
suggest that the enhancement and multiplicity dependence observed in
the inclusive $\Lambda/K^{0}_{s}$ ratio does not result from a
modification of the particle composition within the jets but rather is due to a change in the underlying event or soft processes. 

\section{Conclusions}

The fully reconstructed jet spectra for $R$ = 0.2 and $R$ = 0.4 have been
measured by ALICE in 5.02 TeV p-Pb collisions.  Comparisons to model
predictions of the 5.02 TeV pp jet spectra indicate that the strong
suppression observed in Pb-Pb collisions is a QGP effect and
not an initial state effect. To better quantify the CNM effects, if
any, on the jet spectrum, systematic uncertainties on this measurement must be reduced. In
particular, the uncertainty on the reference can be reduced by measuring
pp collisions at 5.02 TeV. 

The ratio between spectra
reconstructed with different $R$ is consistent with the same ratio in pp
collisions. This also indicates no modification to the substructure of the
jets produced in p-Pb collisions. The fragmentation variable, $j_{T}$,
was also measured. The agreement between the $j_{T}$ distribution shape
measured in p-Pb collisions and the expectations from PYTHIA also
indicate no modification. The data agree best with PYTHIA when
angular ordering is included. 

Measurements of the $\Lambda/K^{0}_{s}$ ratio within a jet show little dependence on
the resolution parameter of the jet or the multiplicity of the event
and are all consistent with PYTHIA calculations. This suggests that the enhancement
of the inclusive $\Lambda/K^{0}_{s}$ ratio for high multiplicity compared to low
multiplicity p-Pb events is due to the underlying event or soft
processes.  All results presented in this talk suggest that there is no CNM
effects to the production, fragmentation or particle composition
of jets in p-Pb collisions.



\begin{thebibliography}{99}


\bibitem{alicePbPbjets} 
R. Reed (for the ALICE Collaboration), Journal of
Physics: Conference Series 446 (2013) 012006 arXiv:1304.5945.

\bibitem{lamkPbPb} ALICE Collaboration, B. Abelev, et al., Phys. Rev. Lett. 111 (2013) 222301, arXiv:1307.5530. 

\bibitem{lamkpPb} ALICE Collaboration, B. Abelev, et al., Phys. Lett. B728 (2014)
25–38. arXiv:1307.6796. 

\bibitem{aliceppjets} ALICE Collaboration, B. Abelev, et al.,
  Phys. Lett. B 722 (2013) 262-272, arXiv:1301.3475. 

\bibitem{fastjet}
M. Cacciari, G. P. Salam, G. Soyez, 
Eur. Phys. J. C72 (2012) 1896, arXiv:1111.6097. 
012- 1896- 2.

\bibitem{cmsrho}
 CMS Collaboration, S. Chatrchyan et al., JHEP08 (2012) 130, arXiv:hep-ex/1207.2392.







\bibitem{V0QM} 
X. Zhang (for the ALICE Collaboration), arXiv:1408.2672.

\bibitem{myQM} 
M. Connors (for the ALICE Collaboration), arXiv:1409.3468.

\bibitem{svd}
A. H\"ocker, V. Kartvelishvili, 
NIM A372 (1996) 469-481, arXiv:hep-ph/9509307. 

\end{thebibliography}
\end{document}